\documentclass[12pt,preprint]{aastex}
\usepackage{amssymb}
\usepackage{pifont}
\usepackage{graphicx}
\RequirePackage{longtable}

\shorttitle{X-ray profile of the Crab Pulsar}
\shortauthors{M.Y. Ge et al.}

\begin{document}

\title{Evolution of the X-ray Profile of the Crab Pulsar}

\author{M.Y. Ge\altaffilmark{1}, L.L. Yan\altaffilmark{1},
F.J. Lu\altaffilmark{1}, S.J. Zheng\altaffilmark{1},
J.P. Yuan\altaffilmark{2}, H. Tong\altaffilmark{2},
S.N. Zhang\altaffilmark{1}, Y. Lu\altaffilmark{3},
}

\affil{$^1$Key Laboratory for Particle Astrophysics, Institute of
High Energy Physics, Chinese Academy of Sciences, Beijing 100049,
 China; gemy@mail.ihep.ac.cn}

\affil{$^2$ Xinjiang Astronomical Observatory, Chinese Academy of
Sciences, Urumqi, Xinjiang 830011, China}

\affil{$^3$ Bureau of Frontier Sciences and Eduation, Chinese Academy of
Sciences, Beijing 100864, China}

\begin{abstract}

Using the archive data from the Rossi X-ray Timing Explorer ({\sl RXTE}),
we have studied the evolution of the X-ray profile of the
Crab pulsar in a time span of 11 years. The X-ray profiles,
as characterized by a few parameters, changed slightly
but significantly in these years: the separation of the two
peaks increased with a rate $0.88\pm0.20\,\textordmasculine$\,per
century, the flux ratio of the second pulse to the first pulse decreased with
$(3.64\pm0.86)\times10^{-2}$\,per century, and the pulse widths of
the two pulses descended with $1.44\pm0.15\,\textordmasculine$\, and
$1.09\pm0.73\,\textordmasculine$\,per century, respectively. The evolutionary
trends of the above parameters are similar to the radio results, but
the values are different. We briefly discussed the constraints of these X-ray
properties on the  geometry of the emission region of this pulsar.

\end{abstract}


\section{Introduction}

The Crab pulsar is one of the most widely studied celestial objects.
It was born in 1054, has a spin period of 33\,ms, and is bright over
the full electromagnetic spectrum from radio to high energy $\gamma$-rays.
At all wavelengths this pulsar shows a double-pulse structure, with the
main pulse (P1) and the inter pulse (P2) separated by a phase of $\sim 144\,\textordmasculine$,
and the exact pulse morphology varies as a function of photon energy
\citep{Eikenberry and Fazio(1997),Kuiper et al.(2001),Rots et al.(1998),
Molkov et al.(2010),Ge et al.(2012)}. Compared to the radio profile,
the X-ray profile has broader pulses and bridge emission that is missing
in the radio band\citep{Eikenberry and Fazio(1997), Massaro et al.(2006)}.
In order to compare the X-ray and radio pulses, the two X-ray peaks
are called P1 and P2\citep{Kuiper et al.(2001)}, which correspond to the radio
main peak (MP) and the inter peak (IP) in \citep{Lyne et al.(2013)}.
The X-ray P1 and the radio MP are not exactly aligned, and the X-ray phase
lag also varies versus energy with the maximum of $-3.6\,\textordmasculine$ at
5$-$9\,keV \citep{Rots et al.(2004),Molkov et al.(2010),Ge et al.(2012)}

In comparison with the phase-resolved spectrum, the long term change of the
pulse profile of the Crab pulsar is relatively less studied.
Nevertheless, it is equally important. The pulse profile is related to
the shape of the emission beams that are determined by the magnetic field structure
\citep{Gold(1968)}. Given the fact that the rotation powered pulsars are very
stable, data with high statistics, frequent and long time span coverage are
needed in order to reveal the very small profile evolution. The Crab pulsar is the best
source for this kind of study, because it is bright at multi-wavelengths, and,
as a calibration source, it is observed very frequently. Previous
studies suggested that the separation of the X-ray P2
and P1 increases with a rate $0.71\pm0.24\,\textordmasculine$\,per century
\citep{Ge et al.(2012)}. Significant radio profile evolution was
discovered by \cite{Lyne et al.(2013)} with the high precision daily observations at
Jodrell Banks Observatory. The separation of the two radio pulses shows
steady increase with $0.62\pm0.03\,\textordmasculine$\,
per century in the past 22 years \citep{Lyne et al.(2013)},
while the relative integrated flux densities
of the two pulses decreased with $-0.172\pm0.008$ per century.
Since the evolution of the pulse separation in the X-ray band and radio
band are similar to each other, although the former is only marginally
detected, the possible X-ray profile evolution may be due to the same
mechanism. Progressive change of the magnetic inclination was used to
explain the radio results \citep{Lyne et al.(2013),Arzamasskiy et al.(2015),
Zanazzi et al.(2015)}, which could also explain the X-ray properties.

Study of the X-ray profile evolution can add more constraints on
pulsar physics. In pulsar emission models, the radio
emission region of a pulsar is thought to be located over the two
magnetic poles, while the high energy emission comes from two
high-altitude gaps that are close to the light cylinder in the
magnetosphere \citep{Cheng et al.(1986a), Cheng et al.(1986b),
Sturner and Dermer(1994),Daugherty and Harding(1996)}. If the
magnetic inclination changes, the X-ray profile should exhibit
simultaneous variation with the radio profile. However, the
changing rates at the two wavelengths might be not the same, because
of the different projection effects along the line of sight. The
evolution of the X-ray and radio profiles of the Crab pulsar
hence presents the three-dimensional information of the magnetosphere
structure with different view angles.

Time related evolution of the pulsar X-ray profile has not been firmly
detected yet, though several studies have been carried out.
\cite{Patt et al.(1999)} reported that the flux densities of the pulses
of the Crab pulsar are steady at the level of 7\% with one hour data
from {\sl RXTE}. \cite{Rots et al.(2004)} and \cite{Chetana et al.(2011)}
also studied the X-ray profile evolution by fitting the profiles and
comparing the fitted parameters, and no evolution was found. As mentioned
previously, \cite{Ge et al.(2012)} found that the separation of the X-ray
P1 and P2 increased at 3\,$\sigma$ level, which is the only
evidence for X-ray profile evolution so far. In this paper, we study the
X-ray profile evolution of the Crab pulsar in details,
by analyzing pulse shape and shape parameters such as pulse
separation, pulse flux ratio and pulse widths, using all the
available archival data from {\sl RXTE}. In order to
make sure that the obtained profile evolution is intrinsic, we also
investigate the variation of the observed profile induced by the
instrumental aging effect.

\section{Observations and Data Reduction}

The {\sl RXTE} data analyzed in this study were obtained by both
the Proportional Counter Array (PCA) and the High Energy X-ray
Timing Experiment (HEXTE). The total exposure time of these
observations is about 230\,ks for PCA and 232\,ks for HEXTE, as
listed in Table \ref{table:ListObsID}. The data reduction was done using
{\sl ftools} from the High Energy Astrophysics Software (HEASoft,
version v6.15).

\subsection{PCA data reduction}

PCA is composed of five Proportional Counter Units (PCUs, named PCU0, PCU1, PCU2, PCU3, PCU4),
which has an effective energy coverage of 2 to 60\,keV, a total
collection area of 6500\,cm$^2$, and the best ever time resolution
of up to 1\,$\mu$s (in Good Xenon mode)\citep{Jahoda et al.(2006)}.
However, data of PCU1 have not been used in this study,
because of its propane loss around 2006-12-25 \footnote{http://heasarc.
gsfc.nasa.gov/docs/xte/pca\_history.html} \citep{Garcia et al.(2014)},
which resulted in a big change of the effective area response and
such a change could distort the evolution of the pulse profile that
varies with energy. Observations with the data mode
E\_250us\_128M\_0\_1s (E\_250us)(Table \ref{table:ListObsID}, MJD
51956--55927) were selected for the final analysis, {\sl because
most of the observations were done in this data mode}. The time
resolution of E\_250us data is about 250\,$\mu$s with all absolute
channels, which were merged to 128 relative channels. Compared to the
dataset used in \cite{Ge et al.(2012)}, about 3 more years of data
were added in the analysis, including ObsIDs P95802, P96382 and P96802, as
listed in Table \ref{table:ListObsID}.

To get the profile of the Crab pulsar from an observation,
we first select the good events and then fold them by using the
{\sl ftools} command FASEBIN and FBSSUM.  The procedure for event
selection is described as follows: (1) create the filter file for
this observation with XTEFILT; (2) generate the Good Time Interval
(GTI) file by MAKETIME using the filter file; (3) using the GTI
file to create the ``good'' events file with GROSSTIMEFILT;
(4) remove the clock events and select the events of PCU0234 from
the event mode data by SEFITER and FSELECT. This procedure is
the same as in Ge et al. (2012). From the data prepared,
phase-resolved spectrum was first created by FASEBIN of {\sl ftools}
with 1000 bins and the JPL ephemeris DE-200 \citep{Standish(1990)}.
This spectrum contains the information of photon count numbers
in each energy channel within the corresponding phase bin.
As there are precise radio ephemerides (including periods and
phases) for the Crab pulsar from Jodrell Bank\footnote{http://www.jb.
man.ac.uk/pulsar/crab.html} \citep{Lyne et al.(1993)}, they were used
in the process to produce the phase-resolved spectrum.
The pulse profile of each observation was obtained from the
phase-resolved spectrum by FBSSUM, which sums up the photon count
numbers in all the energy channels.

The integrated profile in a period was obtained by adding all the
individual profiles from the observations included. Phase 0 of a profile
was obtained by a cross-correlation analysis between this profile and the
reference one. When we produce the total profile from all the observations
in the 11 years, the reference profile is the one generated from the
first observation. But when we produce the integrated profile in a
time period, the reference profile is actually the total profile.

\subsection{HEXTE data reduction}

The HEXTE instrument consists of two independent detector
clusters A and B, each containing four Na(Tl)/CsI(Na) scintillation
detectors \citep{Rothschild et al.(1998)}. The HEXTE detectors
are mechanically collimated to a $1^{\,\textordmasculine}$ field of
view and cover the 15-250\,keV energy range with an collecting
area of 1400\,cm$^{2}$. In its default operation mode, when one cluster
points to the target, the other one is off from the source to provide
instantaneous background measurements. Because of the co-alignment of the HEXTE
and the PCA, the two instruments observe the same target simultaneously.
In this analysis, data from both Cluster A and B were used before 2009-12-14,
and after that date only Cluster A data are available because cluster B ceased
modulation and stared off-source position thereafter\footnote{http://heasarc.gsfc.
nasa.gov/docs/xte/whatsnew/newsarchive\_2010.html\#hexteB\_locked}.

The data mode selected for the pulse profile was E\_8us\_256\_DX0F that with
a time resolution of 8\,$\mu$s. The standard data reduction was performed
and the data were extracted from clusters A and B separately. Pulse profiles
were obtained using FASEBIN and FBSSUM of {\sl ftools}, with the same
timing parameters and procedure as did for the PCA profiles.

\section{Evolution of the X-ray profile}

In this section, the evolution of the X-ray profile will be studied
in two ways, using the profile ratio curve and the profile parameters respectively.
The profile ratio is the ratio between two normalized profiles in
different epochs, which can exhibit the profile variation directly
if the profile changes with time. On the other hand, in order to
present the profile evolution in a more quantitative way, we have
used four parameters to characterize the X-ray profiles in different
epochs and to study the changes of these parameters versus time.

\subsection{Profile ratio from PCA}

\begin{figure}
\begin{center}
\includegraphics[width=0.8\textwidth]{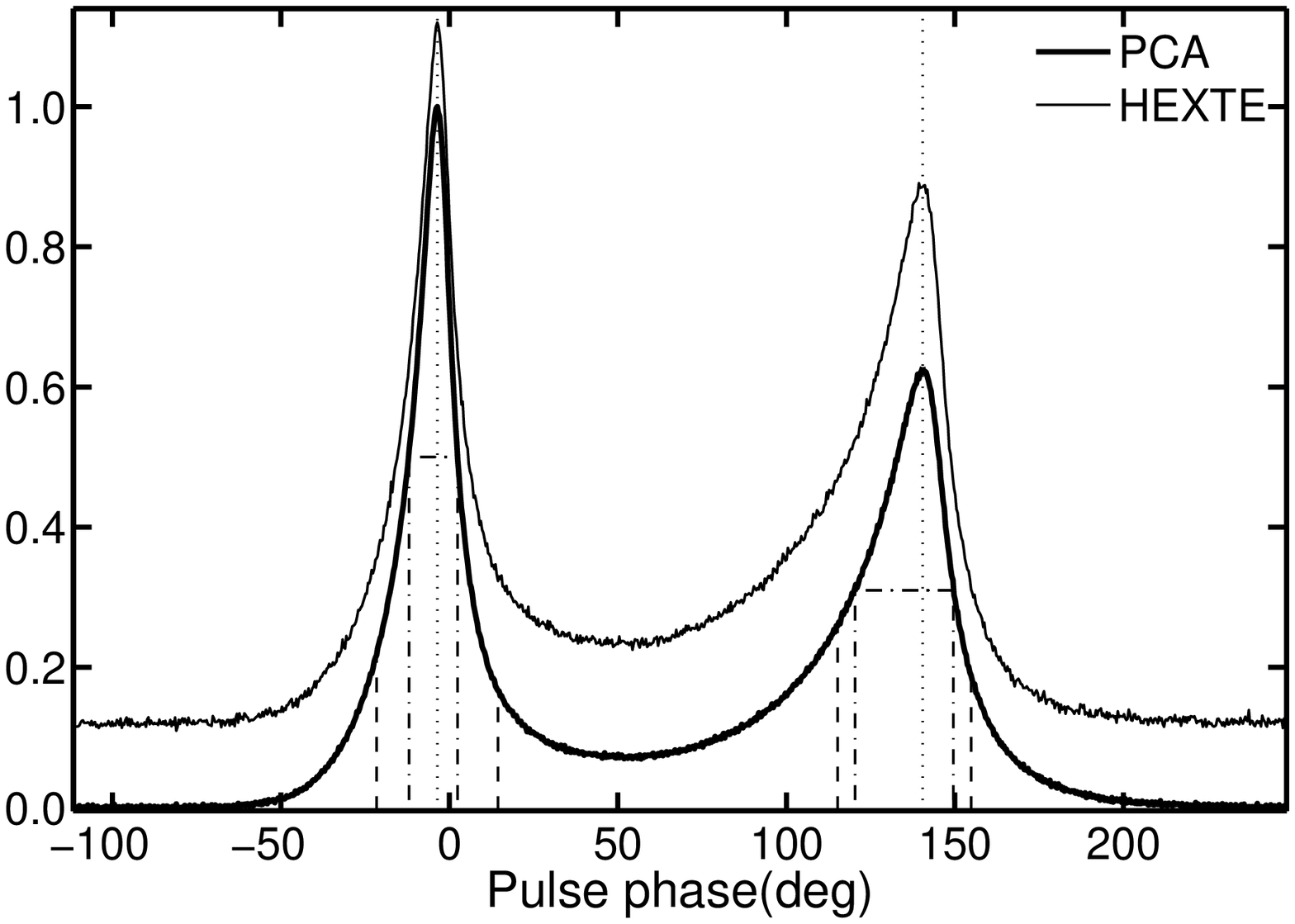}
\caption{
The total normalized profiles of the Crab pulsar measured by {\sl RXTE}
(PCA, MJD 51956--55927, 2--60\,keV: thick line, HEXTE, MJD
51302--55927, 15--250\,keV: thin line). The profiles were
shifted with 0.0 and 0.12, respectively.
Phase 0 represents the phase of the radio pulse. The two dotted lines
represent the peak positions of the two pulses. The two dotted-dashed lines
around P1 denote the phase range in which the flux was integrated, and the
two dotted-dashed lines around P2 have the same meaning.
The dashed lines are similar to the dotted-dashed lines but
represent the phase ranges defined in \cite{Kuiper et al.(2001)}.
\label{fig}}
\end{center}
\end{figure}

\begin{figure}
\begin{center}
\includegraphics[width=0.8\textwidth]{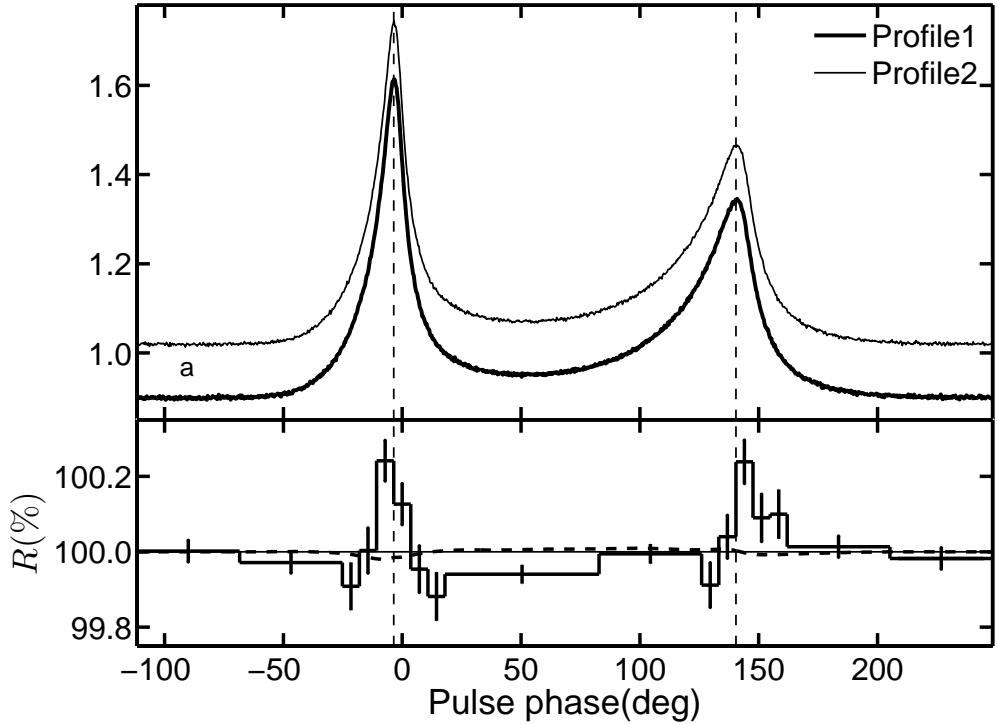}
\caption{
The two X-ray profiles of the Crab pulsar obtained by PCA in different time periods
and their ratio curves.
Panel $a$ shows Profile1 (thick line, integrated in MJD 51956 to 52500) and
Profile2 (thin line, in MJD 55254 to 55927, shifted by a value of 0.12). The solid
line in panel $b$ shows the Profile2 to Profile1 ratio, and the dashed line
represents the faked ratio curve induced by the instrument aging as described in
section 4.1. The two vertical lines denote the peak
positions of the two pulses.
\label{fig0}}
\end{center}
\end{figure}

If the X-ray profile of the Crab pulsar showed detectable secular
changes, the ratio curve between the normalized profiles in
different epochs should deviate from a uniform distribution.
The total profiles obtained from the 11-year observations
of PCU0234 and HEXTE are shown in Figure \ref{fig}, in which the
PCA profile shows a much higher statistics than the HEXTE one.
Therefore, two groups of profiles obtained from PCA data that
obtained in time range of MJD 51955(2001-02-15)--52500(2002-08-14) and
MJD 55254(2010-02-27)--55927(2012-01-01), were created, and then
merged into two integrated profiles named Profile1 and Profile2, respectively.
These two integrated profiles are used to produce the ratio curve.
In order to eliminate the effect of the different background
levels in these two periods, the two integrated profiles were
reformed as follows: (1) subtract the background level that is
determined by the mean flux per bin in phase 226.8
to $298.8\,\textordmasculine$ as \cite{Ge et al.(2012)};
(2) normalize the pulse profile to make the integrated flux, which is
the sum of the flux per bin times the phase bin size, equal to 1;
(3) add 9.0 to the value in each bin, because the total background
count rate is about 9 times as high as the count rate of the pulses; and
(4) normalize the pulse profile again. After these steps, the Profile2
to Profile1 ratio was obtained as shown in Figure \ref{fig0}b.

As can be seen clearly in Figure \ref{fig0}b, the distribution
of the ratio is not uniform, with the $\chi^{2}$ of 63.2
for 17 points. The ratio at P1 (phase: -7.2 to $7.2\,\textordmasculine)$
is higher than 1.0 with the mean of 1.0016, and lower than 1.0 with
mean of 0.9994 at the bridge(50.4--$90\,\textordmasculine$,
as defined in \cite{Kuiper et al.(2001)}). Compared to P1, P2
shows different behaviors that the ratio is around 1.0 at the
leading edge (phase: 126--$144\,\textordmasculine$) and is higher
than 1.0 at the trailing edge (phase: 144--$158\,\textordmasculine$).
The deviation of the profile ratio curve from a uniform distribution
implies that the X-ray profile evolved with time: after a few years,
P1 became sharper and the distance between P1 and P2 increased a
little bit.

\subsection{Parameterization of the X-ray profile}

Since the X-ray profile of the Crab pulsar has a typical double-peak structure,
we used four parameters, including separation ($\Phi$), flux ratio ($R_{f}$),
and widths ($W_{1}$ $\&$ $W_{2}$) of P1 and P2, to quantify the X-ray profile. $\Phi$ is the
relative phase distance between the two maxima of the pulses. $W_{1}$ and $W_{2}$ are
defined as the full width at half maximum (FWHM) of P1 and P2 after subtracting
the pulse background (phase 226.8--$298.8\,\textordmasculine$). $R_{f}$
is the ratio between the integrated fluxes of P1 (-12.24 to
$2.52\,\textordmasculine$) and P2 (123.84--$153.36\,\textordmasculine$),
i.e., the integrated flux within the FWHM of the two pulses.

The fluxes of P1 and P2 (as well as their errors) could be obtained directly,
but more calculations are needed for $\Phi$, $W_{1}$ and $W_{2}$ to have accuracy finer than
the bin size. To obtain the accurate peak position and width of a pulse, we fitted
it using the empirical formula proposed by \cite{Nelson et al.(1970)},
\begin{equation}
L(\phi-\phi_{0})=N\frac{1+a(\phi-\phi_{0})+b(\phi-\phi_{0})^{2}}
{1+c(\phi-\phi_{0})+d(\phi-\phi_{0})^{2}}e^{-f*(\phi-\phi_{0})^{2}}+l
\label{eq:1}
\end{equation}
where $L$ is the intensity at phase $\phi$, $l$ the baseline
of the pulse profile, $\phi_0$ the phase shift, $N$ the pulse
height of the profile, and $a$, $b$, $c$, $d$ and $f$
the shape coefficients. The pulse phase is measured in degrees.
In the fitting, P1 was chosen in the phase range -27 to
$12.78\,\textordmasculine$ and P2 in 109.8 to $167.4\,\textordmasculine$.

The peak separation $\Phi$ is calculated with the
following steps(similar with \cite{Ge et al.(2012)}):
(1) Fit the two pulses of the total profile with Nelson's formula
and obtain their shape coefficients $a$, $b$, $c$, $d$, $f$, as
well as $N$, $l$, and $\phi_{0}$, which are listed in
Table. \ref{table:nelsonpar};
(2) For an integrated profile in a time period and from which
we want to get $\Phi$, fit its two pulses using the Nelson
formula with $N$, $\phi_{0}$, and $l$ free and the
other shape coefficients fixed to the values in
Table. \ref{table:nelsonpar};
(3) From the positions of the two maxima of the fitted profiles get their
separation.

$W_{1}$ and $W_{2}$ were obtained in a similar way. We fitted the
observed profile with the Nelson's formula with all the coefficients
free, and then the FWHMs of the fitted profiles were taken as the
widths of P1 and P2.

We estimated the errors of $\Phi$, $W_{1}$ and $W_{2}$ with a
Monte Carlo method. 100 simulated profiles were created by sampling
from the original profile under the assumption that the photon counts
in every phase bin follow the Poisson distribution. Then these simulated
profiles were fitted with Nelson's formula and 100 groups of $\Phi$,
$W_{1}$, and $W_{2}$ values were obtained. The distribution widths
of these values can represent their statistical errors, and the
1$\sigma$ width of the Gaussian function fit to the distribution
of a parameter was taken as its 1$\sigma$ error.

\subsection{Profile evolution with time}

To study the evolution of $\Phi$, $R_f$, $W_1$ and $W_2$,
we divide the PCA observations into 6 groups with roughly equal
time span of about 660 days. In one group an integrated profile
was produced and a set of $\Phi$, $R_f$, $W_1$ and $W_2$ were
derived, so we have 6 data points for every parameter from the
PCA observations. Because the source photons collected by
HEXTE are much less than that by PCA, two profiles and thus two
set of data points from HEXTE data were obtained in time
ranges MJD 51302(1999-03-08)--54789(2008-11-19) and
MJD 52570(2002-10-23)--55927(2012-01-01), respectively.

\begin{figure}
\begin{center}
\includegraphics[width=0.8\textwidth]{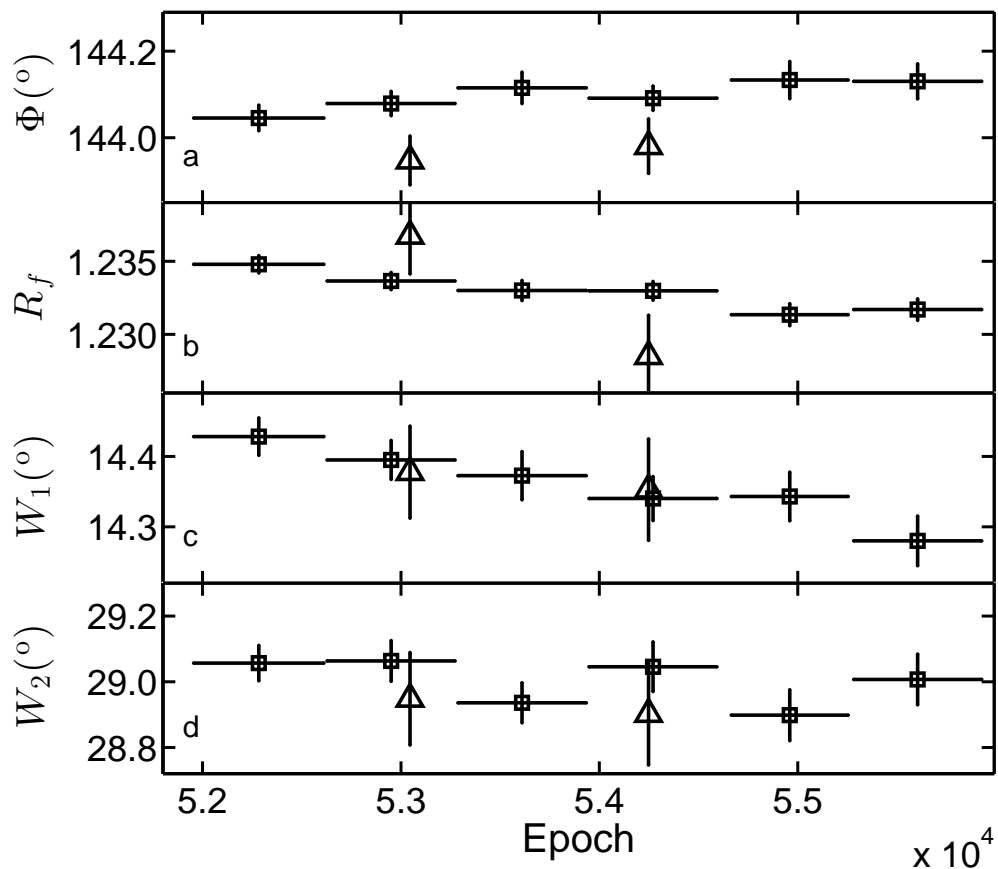}
\caption{
The profile parameters of the Crab pulsar from PCA (squares) and
HEXTE (triangles) observations. Panels $a$, $b$, $c$ and $d$
show the peak separation ($\Phi$), flux ratio ($R_{f}$),
and widths ($W_{1}$ $\&$ $W_{2}$) of P1 and P2, respectively.
The data points of ${R}_{f}$, ${W_{1}}$ and ${W_{2}}$ from
HEXTE data were shifted by $-0.273$, $-0.5\,\textordmasculine$
and $-0.7\,\textordmasculine$, respectively.
\label{fig1}}
\end{center}
\end{figure}

Figure \ref{fig1} displays  $\Phi$, $R_{f}$, $W_{1}$, and
$W_{2}$ of the Crab pulsar measured from PCA and HEXTE. The
parameters of the PCA profiles show large differences from the
HEXTE ones, because the X-ray profile varies with energy
\citep{Mineo et al.(1997),Massaro et al.(2006)}.
However, the variation trends of the parameters
from these two instruments are similar. The peak
separation $\Phi$ increased with time and the other
parameters decreased with time.

\subsubsection{Profile evolution results from the PCA data}

The PCA data were first chosen to study the profile evolution
using the four parameters defined above, which were fitted
with linear functions to derive their changing rates.
As listed in Table \ref{table:para_slope}, the evolutions of
$R_f$ and $W_{1}$ were detected with a high significance:
the changing rate of $R_{f}$ is $-(3.52\pm0.64)\times10^{-2}$\,per
century, and that of $W_{1}$ is $1.45\pm0.19\,\textordmasculine$\,per
century. The evolutions of $\Phi$ and $W_{2}$ are less significant,
with $0.89\pm0.26\,\textordmasculine$\,  and
$-1.09\pm0.95\,\textordmasculine$\,per century, respectively.
The changing rate of the peak separation is consistent
with the result from \citep{Ge et al.(2012)} within 1$\sigma$
error, and the new result has a little bit higher significance.
These quantitative results for the peak separation and flux
ratio are consistent with the qualitative results deduced
from the profile ratio curve.

\begin{figure}
\begin{center}
\includegraphics[width=0.8\textwidth]{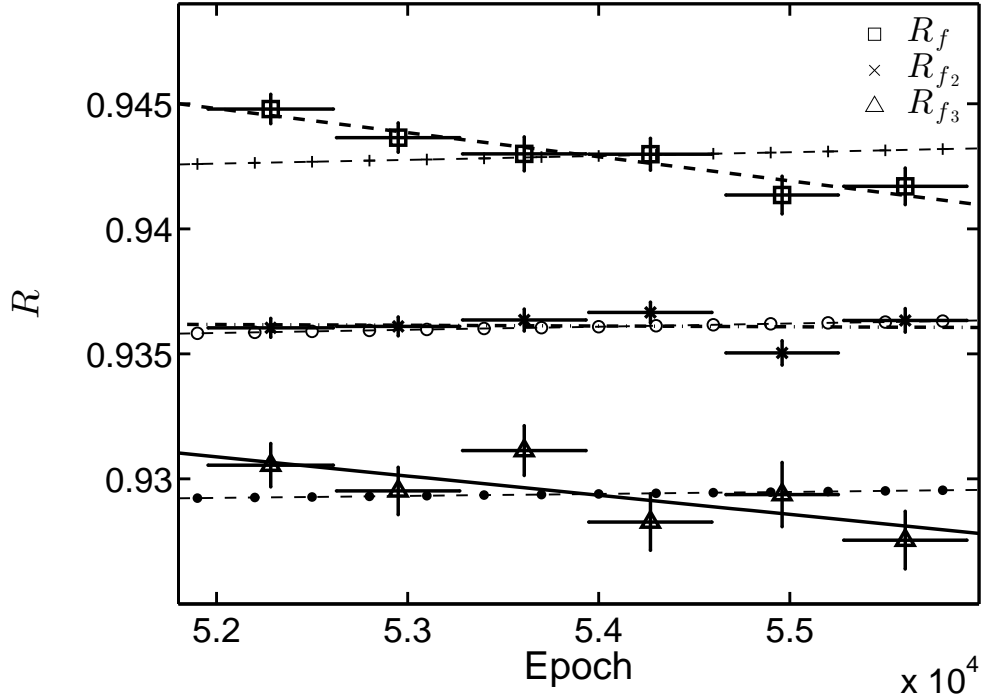}
\caption{
The pulse flux ratios of the Crab pulsar calculated in
different phase ranges for PCA. The square points represent the
ratios  of the fluxes integrated in the FWHM of the two pulses
($R_{f}$) with $-0.29$ shifted, the cross points represent the
ratios of the fluxes integrated in $-21.6$ to 14.4\,\textordmasculine
(for P1) and 115.2 to 154.8\,\textordmasculine (for P2) ($R_{f_2}$),
and the triangle points represent the ratios of the fluxes at the
two peaks ($R_{f_3}$, shifted by 0.31). The dashed, dotted-dashed
and solid lines are the fitted results for $R_{f}$, $R_{f_2}$ and
$R_{f_3}$, respectively. Similarly, the thin dashed lines with ``+'', empty
and filled circles are the fitted results for $R_{f}$, $R_{f_2}$ and
$R_{f_3}$ obtained from the faked profiles as described in Section 4.1.
The phase ranges for the flux ratio calculation are shown in Figure \ref{fig}.
\label{fig2_2}}
\end{center}
\end{figure}

The flux ratio of the two pulses should be different if
the fluxes were integrated in different phase ranges. In order to
get more information of the profile evolution, two more flux
ratios were calculated with different phase ranges as shown
in Figure \ref{fig}. $R_{f_2}$ represents the integrated flux ratio of
{\sl P1} (-21.6 to $14.4\,\textordmasculine$) and {\sl P2}
(115.2--$154.8\,\textordmasculine$) (as defined in
\cite{Kuiper et al.(2001)}). The mean value of $R_{f_2}$
is 0.936(5), which is consistent with the result of
\cite{Kuiper et al.(2001)}. $R_{f_3}$ is the peak flux ratio
of P1 and P2, which has a mean value 0.619(1). As shown in
Figure \ref{fig2_2}, the changing rates of $R_{f_2}$ and
$R_{f_3}$ are $(-0.1\pm0.8)\times10^{-2}$\,per century
and $(-2.8\pm1.5)\times10^{-2}$\,per century,
respectively. Apparently, no evolution has been detected for
$R_{f_2}$, while the changing rate of $R_{f_3}$ is close
to $R_{f}$ but with a much lower significance.
In any case, the changing rates of these two flux
ratios are also significantly smaller than the radio
results.

\subsubsection{Joint study of the profile evolution with PCA and HEXTE}

\begin{figure}
\begin{center}
\includegraphics[width=0.8\textwidth]{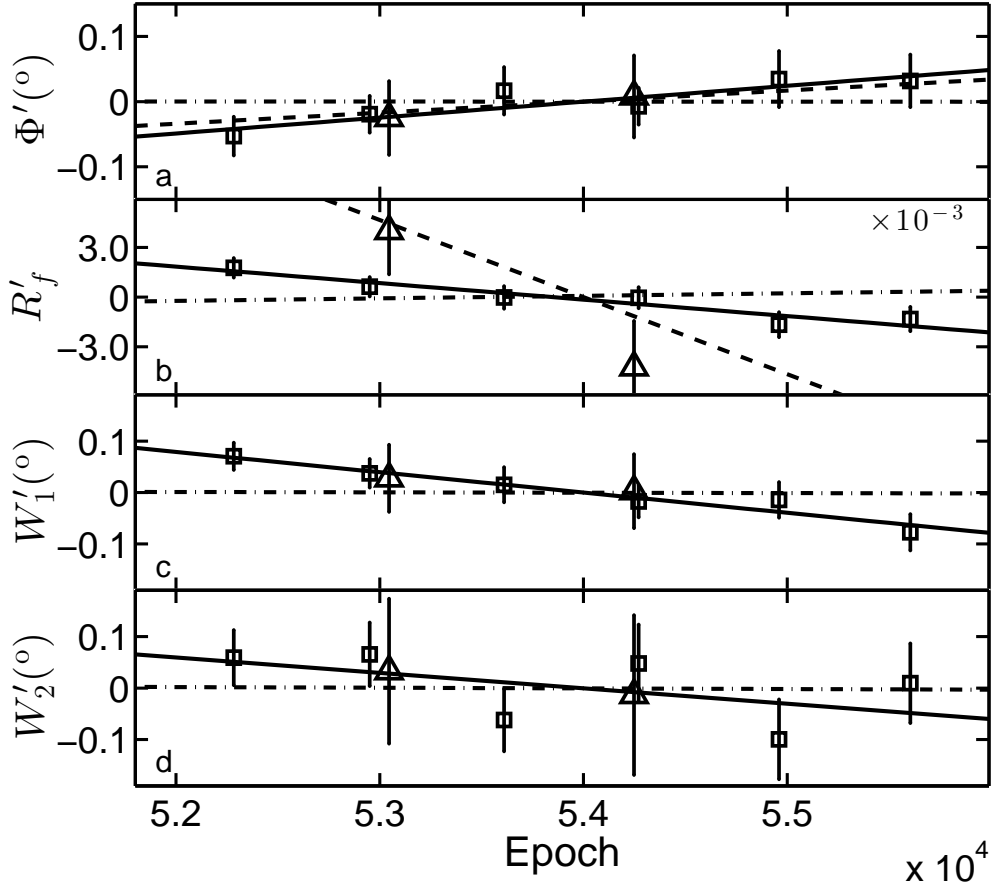}
\caption{
The corrected profile parameters from PCA (squares) and HEXTE (triangles).
Panels $a$, $b$, $c$ and $d$ show the corrected parameters $\Phi'$,
$R_{f}'$, $W_{1}'$ and $W_{2}'$, respectively. The dashed lines
in panels $a$ and $b$ represent the radio results \citep{Lyne et al.(2013)}, and
the dotted-dashed lines are the fitted results of the parameters from the
faked profile as shown in Figure \ref{fig3}.
\label{fig2}}
\end{center}
\end{figure}

Figure \ref{fig1} shows that the parameters infered from the HEXTE data show
evolution similar to the PCA results, even though the parameters from two kinds
of instruments have different mean values. With the assumption that the
secular changes of the profile measured from PCA and HEXTE follow the same
trends, the PCA and HEXTE parameters were shifted to around zero so that
the $\chi^2$ of the linear fitting, which is defined by equation \ref{eq:2},
reaches the minimum.
\begin{equation}
\chi^2 = \frac{1}{N-3}\sum_{ij}(\frac{k(t_{ij}-t_{0})+b_{i}-\Phi_{ij}}{\sigma_{ij}})^{2}
\label{eq:2}
\end{equation}
where $N$ is the number of points, $k$ is the changing rate, $i$ is 0 or 1 to represent
the data from PCA or HEXTE, $j$ denotes the data points from each instrument,
$\Phi_{ij}$ is the peak separation, and $\sigma_{ij}$ the error of
the peak separation obtained with the Monte-Carlo method described in section 3.1.
The intercept $b_{i}$ correspond to the values at $t_0=54000$ in MJD format.
The best estimation for $b_{i}$ was obtained when $\chi^{2}$
reached the minimum. Then, the corrected peak separation $\Phi'$
was obtained with $b_{0}=144.10\,\textordmasculine$ and
$b_{1}=143.98\,\textordmasculine$ subtracted. As presented in
Figure \ref{fig2}, $\Phi'$  increased linearly with time.
$R_f$, $W_1$ and $W_2$ were processed with the same method, and the
shifted intercepts for all these parameters are listed in Table
\ref{table:para_shift}.

$\Phi$, $R_f$, and $W_{1}$ inferred from the joint PCA and HEXTE data show secular
changes similar to that from PCA data alone (Figure \ref{fig2}). The evolution
of $W_{2}$ has not been significantly detected either, even with the HEXTE data added.
The changing rate of the peak separation is $0.88\pm0.20\,\textordmasculine$\,per
century, which is similar to the radio result, $0.62\pm0.03\,\textordmasculine$\,per
century \citep{Lyne et al.(2013)}. Ratio of the integrated fluxes within the
FWHM of the two pulses descended with a rate $(3.64\pm0.86)\times10^{-2}$\,per century,
about 1/5 of the radio result, $(17.2\pm0.8)\times10^{-2}$\,per
century \citep{Lyne et al.(2013)}. $W_1$ and $W_2$ descend with
slopes of $1.44\pm0.15\,\textordmasculine$\,per century and
$1.09\pm0.73\,\textordmasculine$\,per century, respectively.

\section{Discussion}
In this section, we will first study whether the observed profile evolution is
due to the aging of the instruments, and then discuss the possible constraints
of our results on the geometry of pulsar's magnetosphere.

\subsection{Profile changes induced by the aging of PCA}

Previous studies have shown that the X-ray profile of the Crab pulsar varies with
photon energy \citep{Eikenberry and Fazio(1997),Rots et al.(1998),
Massaro et al.(2000), Willingale et al.(2001),
Molkov et al.(2010),Ge et al.(2012)}, which can be also described
quantitatively by the phase-resolved spectrum, and this has been studied in details by
\citep{Ge et al.(2012)} using the PCA observations (from 2001-02-15 to 2009-11-07).
Because the X-ray profiles we studied in this paper were integrated over the entire
energy bands of the instruments, if the effective area of the instrument at different
energies had gradual changes in the about 11 years of operation \citep{Garcia et al.(2014)},
it could also result in pseudo changes of the X-ray profiles. In order to study
the influence of the instrument aging on the observed profile evolution, we use
the phase-resolved spectrum $F(\phi,E)$ measured from PCA \citep{Ge et al.(2012)}
as input, convolve it with the effective area curves of PCU0234 in different
epochs to fake the profiles, and from these profiles further derive the shape
parameters in those epochs. Comparison of the shape parameters of the faked
profiles and those of the observed ones will verify whether the profile evolution
we found is intrinsic or not. The detailed process and results are given below.

The input profile $F(\phi,E)$ can be expressed as
\begin{equation}
F(\phi,E)=\beta(\phi)(E/E_{0})^{-\alpha(\phi)}\exp[{-N_{H}\sigma(E)}]
\label{eq:3}
\end{equation}
where the normalized flux $\beta(\phi)$ and photon index $\alpha(\phi)$
were inferred from \cite{Ge et al.(2012)}, $E$ is the photon energy
in units of keV and $E_{0}=1$\,keV, the absorption column density
$N_{H}$ is $0.36\times10^{22}$\,cm$^{-2}$ \citep{Ge et al.(2012)},
$\sigma(E)$ is the photo-electric cross-section
\citep{Morrison and McCammon(1983)}. The number of input data from Table 5
of Ge et al. (2012) are not fine enough to get a smooth pulsed profile,
so the normalized fluxes were fitted with two combined Nelson
formula \citep{Nelson et al.(1970) and four Gaussian functions, which
were mainly used to fit flux in the bridge region,
and the flux in each phase bin was obtained from the interpolation of the fitted
functions}. The photon indices of the pulsed emission in \citep{Ge et al.(2012)}
were first smoothed with a Gauss function that has an 1-sigma width of 3.6 degrees,
and the smoothed photon indices were further fitted with an 18 order polynomial
to get the photon index for a phase bin in this paper. Since there were only
observed photon indices in the phase range greater than -48.6 and smaller
than 196.2 \citep{Ge et al.(2012)} the polynomial function can not be
constrained outside of this phase range. Therefore, the photon indices
(of the pulsed emission) for bins with phases smaller than -48.6 degrees
were fixed as 1.72, the value at -48.6 degrees, and for bins with phases
greater than 192.2 degrees the photon indices were fixed as 2.17.
The pulse profile $F_{P}(\phi)$, which is
the phase resolved spectra (of the pulsar plus nebula) convolved with
the response matrix, was obtained as follows
\begin{equation}
F_{P}(\phi)={\sum_{Ei}{F(\phi,E)RSP(E,i)}}\Delta{E}
\label{eq:4}
\end{equation}
where $RSP(E,i)$ is the average response matrix of PCU0234, and $\Delta{E}$
the input energy width at energy $E$.

The dead time of the instrument could also change the pulse
profile, because the photon fluxes at different phases are
not the same. We therefore calculated the dead time correction
$DCOR$ using the formula:
\begin{equation}
DCOR=\frac{T}{T-DTF}
\label{eq:5}
\end{equation}
\begin{equation}
DTF=C\times{T}\times{dt}
\label{eq:5}
\end{equation}
where $T$ is the length for one phase bin, $C$ the photon
counts from both the pulsar \citep{Ge et al.(2012)} and
nebula \citep{Garcia et al.(2014)} in $T$, $dt$ the time
for {\sl RXTE}/PCA to process the information of one event
\citep{Jahoda et al.(2006)}, which is also called the dead time.
$T$ was 33.6\,$\mu$s because the profile was divided into 1000
phase bins in the calculation. The background events were also
considered when we calculated $DTF$\footnote{http://heasarc.gsfc
.nasa.gov/docs/xte/recipes/pca\_deadtime.html}. The response
profile has been divided by $DCOR$ when we produce the ``final''
faked profiles.

\begin{figure}
\begin{center}
\includegraphics[width=0.8\textwidth]{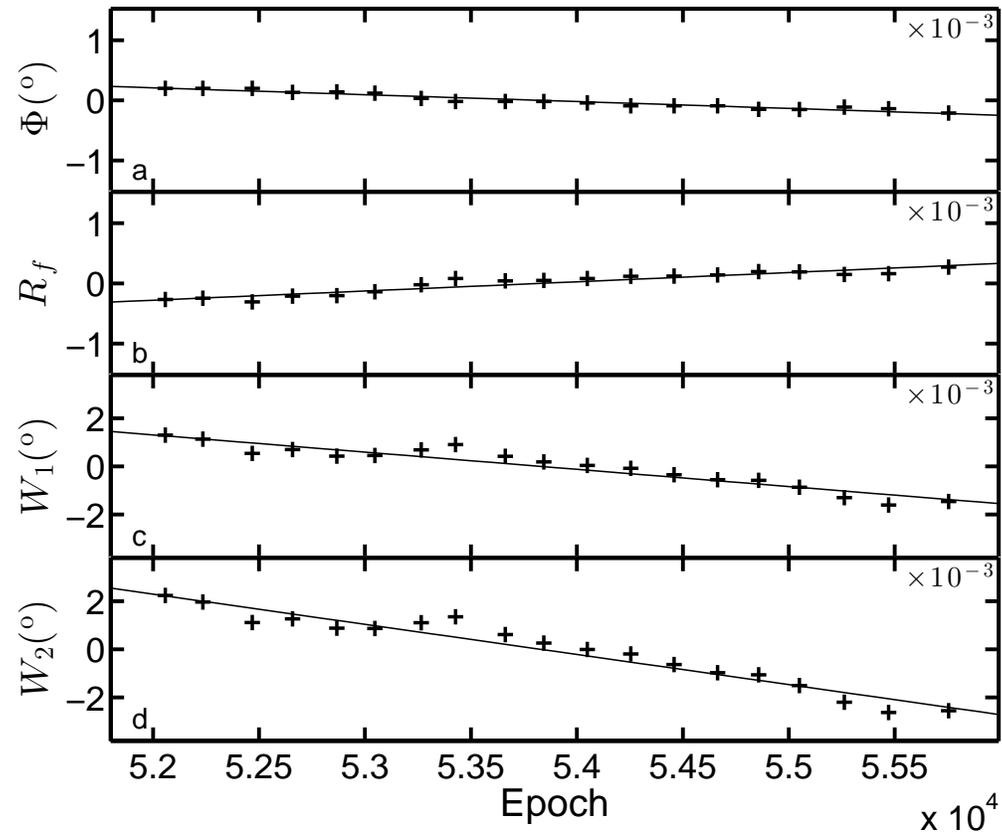}
\caption{
The parameters derived from the faked PCA profiles representing the influence of the
aging of PCA. The solid lines are the linear fits to these parameters.  The mean values
of these parameters were subtracted as we were only interested in their variations.
\label{fig3}}
\end{center}
\end{figure}

Similar to what we did previously, using the faked profiles,
we can obtain the ratio curve and parameter changes that induced
by the instrument aging. The dashed line in the lower panel
of Figure \ref{fig0}b represents the ratio of the faked profiles
in MJD 51956--52500 (Profile3) and MJD 55254--55927 (Profile4).
It is very different from the observed one and with a much smaller
amplitude. The variations of $\Phi$, $R_{f}$, $W_{1}$, and $W_{2}$
induced by the aging of PCA are presented in Figure \ref{fig3},
and the linear fitting results are listed in Table
\ref{table:para_slope}. The changing rates are about 1 to
2 orders of magnitudes lower than the observed ones, and particularly
for $\Phi$ and $R_{f}$, the instrument aging had secular changes
opposite to the observed ones. Similarly, the changing
rates $R_{f2}$ and $R_{f3}$ are $(0.29\pm0.01)\times10^{-2}$\,per century
and $(0.45\pm0.01)\times10^{-2}$\,per century, which also had the
secular changes opposite to the observed ones. Therefore, the
contribution of the instrument aging to the observed pulse profile
evolution is negligible.

\subsection{Constrains on the geometry of the magnetosphere}

The X-ray profile of the Crab pulsar shows secular
changes that the peak separation of the profiles increases
while the flux ratio and widths of the two pulses decrease
with time. The evolutionary trends of the X-ray profile are
similar to the radio results, which means that the
magnetosphere evolution has a similar effect
on emission regions of the X-ray and radio pulses.

For a simple magnetic dipole, the evolution of the magnetosphere
axis is expected
towards alignment rather than orthogonality \citep{Lyne et al.(2013),
Philippov et al.(2014),Arzamasskiy et al.(2015)}. However, the secular
increases of the peak separations in the radio and X-ray bands are inconsistent
with this expectation. \cite{Lyne et al.(2013)} explained the evolution
with geometrical model that inclination of the magnetic axis increases
with time as the torque developed by the return current in the neutron
star surface \citep{Beskin et al.(2007)}. Based on the magnetohydrodynamic
simulations \citep{Philippov et al.(2014)} and \cite{Arzamasskiy et al.(2015)}
pointed out that it is the magnetic dipole precession behavior with
a characteristic time of 100\,yr. The similar evolutionary rates of the X-ray
and radio peak separations imply that the
two kinds of emission locates at similar latitudes.

The evolution in the relative flux densities of the radio components are
explained as highly coherent. Narrow beam and small structural
magnetosphere changes might cause large effects on the component
flux densities \citep{Lyne et al.(2013)}. However, the changing
rate of $R_f$, $(3.64\pm0.86)\times10^{-2}$\,per century, is significantly
lower than the radio result, $(17.2\pm0.8)\times10^{-2}$\,per century. The
difference even became bigger if wider phase intervals were chosen in
calculating the integrated fluxes. This means that the X-ray emitting
region is much larger than the radio emitting region, consistent with
those represented by the widths of the radio and X-ray pulses.
Therefore, a more complicated model is needed to explain the overall
evolutionary behaviors of the radio and X-ray profiles, combined
with the effects of the propagation time and the relativity \citep{Morini(1983)}.

\section{Summary}
In this paper,  we found that the X-ray profile of
the Crab pulsar had secular changes with time. The
ratio curve of the two profiles in different epochs
showed that, after a few years, P1 became sharper
and the distance between P1 and the P2 increased a
little bit. Quantitatively, the peak separation of
the two pulses increased with $0.88\pm0.20\,\textordmasculine$\,per
century, ratio of the integrated flux of P1 to that of P2
decreased with $(3.64\pm0.86)\times10^{-2}$\,per
century, and the widths of the P1 and P2 changed with $-1.44\pm0.15
\,\textordmasculine$\,per century and $-1.09\pm0.73\,
\textordmasculine$\,per century, respectively. These
evolutionary trends are similar to the radio trends,
although the values are different. A more complicated
model of pulsar emission geometry is needed to explain
the radio and X-ray results simultaneously.

\section*{Acknowledgments}

Drs. Michael Smith, Lorenzo Natalucci, Craig Markwardt, Yuanyue Pan,
Lingming Song and Jinlu Qu, are appreciated for their useful
suggestions. We thank the High Energy Astrophysics Science Archive
Research Center (HEASARC) at NASA/Goddard Space Flight Center for
maintaining its online archive service that provided the data used
in this research. This work is supported by National Science
Foundation of China (11233001 and 11503027) and the Strategic
Priority Research Program on Space Science, the Chinese
Academy of Sciences, Grant No. XDA04010300.

\clearpage

\begin{table*}
\footnotesize
\caption {The observation list of {\sl RXTE} used in this work}
\scriptsize{}\label{table:ListObsID}
\medskip
\begin{center}
\begin{tabular}{ c c c c c c c}
\hline \hline

& Obs ID & Start Date & End Date & offset(') & PCA exposure [s] & HEXTE exposure [s] \\
\hline
& 40093 & 1999-03-08 & 1999-03-22 & 0.03 &  --    & 13724 \\
& 50098 & 2000-07-17 & 2000-07-21 & 0.03 &  --    & 5474 \\
& 50099 & 2001-02-15 & 2001-08-27 & 0.03 & 7504   & 15466   \\
& 60079 & 2001-09-10 & 2002-10-22 & 0.03 & 20240  & 18632   \\
& 60080 & 2001-07-18 & 2001-07-20 & 0.03 & 3776   & 3701  \\
& 60420 & 2001-09-07 & 2001-09-09 & 0.03 & 1824   & 1718 \\
& 70018 & 2002-05-09 & 2003-05-14 & 0.03 & 9616   & 7024 \\
& 70802 & 2002-11-07 & 2003-02-26 & 0.03 & 7888   & 7263  \\
& 80802 & 2003-03-13 & 2004-02-15 & 0.03 & 17552  & 17221   \\
& 90129 & 2004-11-15 & 2004-11-18 & 0.03 &   --  & 5946 \\
& 90802 & 2004-02-29 & 2005-02-25 & 0.03 & 18032  & 16086\\
& 91802 & 2005-03-13 & 2006-02-10 & 0.03 & 17088  & 13971\\
& 92018 & 2006-05-10 & 2006-12-21 & 0.03 &   --   & 9774 \\
& 92802 & 2006-03-11 & 2006-09-24 & 0.03 & 23536  & 25704\\
& 93802 & 2007-07-17 & 2008-12-17 & 0.03 & 28768  & 30164\\
& 94802 & 2008-12-31 & 2009-11-07 & 0.03 & 16768  & 13922\\
& 95802 & 2010-01-03 & 2010-12-08 & 0.03 & 29028  & 16229$^{[1]}$\\
& 96382 & 2011-10-17 & 2011-12-11 & 0.03 & 8211   & --      \\
& 96802 & 2011-12-17 & 2012-01-01 & 0.03 & 19920 & 15301$^{[1]}$\\
\hline
& total exposure(s) & & & & 229751  &231649\\
\hline
\end{tabular}
\end{center}
[1] Only cluster A data of HEXTE were used in our analyses.
\end{table*}

\begin{table*}

\footnotesize
\caption {The coefficients of the Nelson's formula for the main peak (P1) and the second peak (P2) of the Crab pulsar}
\scriptsize{}\label{table:nelsonpar}
\medskip
\begin{center}
\begin{tabular}{l c c c c c c c c c}
\hline \hline
&  &             & a & b & c & d & f & $\chi^{2}_{dof}$ & d.o.f.\\
\hline
& PCA   & P1    &  -30.79 & 1550.04 &  -55.03 & 4521.40 &  568.32 & 2.0 & 92 \\
&       & P2    &  -29.84 &  372.30 &  -42.00 & 1265.97 &  138.09 & 2.4 & 128 \\
\hline
& HEXTE & P1    &  -34.26 & 2351.16 &  -49.04 & 5971.22 &  667.86 & 2.0 & 92 \\
&       & P2    &  -26.94 &  209.19 &  -45.95 & 887.59 &  131.46 & 1.7 & 128 \\
\hline
\end{tabular}
\end{center}
\end{table*}

\begin{table*}
\caption {The changing rates of the X-ray profile parameters of the Crab pulsar}
\scriptsize{}
\footnotesize
\label{table:para_slope}
\begin{center}
\begin{tabular}{ l l c c c c c }
\hline
\hline
& Instrument & $\Phi$          & $R_{f}$    & $W_{1}$ & $W_{2}$ \\
&            & (\,\textordmasculine/century)  & ($10^{-2}$) & (\,\textordmasculine/century) & (\,\textordmasculine/century) \\

\hline
& PCA            & $  0.89 \pm  0.26 $ & $ -3.52 \pm  0.64 $ & $ -1.45 \pm  0.19 $ & $ -1.09 \pm  0.95 $\\
& All$^{[1]}$    & $  0.88 \pm  0.20 $ & $ -3.64 \pm  0.86 $ & $ -1.44 \pm  0.15 $ & $ -1.09 \pm  0.73 $\\
& Resp$^{[2]}$   & $ -0.004 \pm  0.001 $ & $  0.56 \pm  0.01 $& $ -0.026 \pm  0.001 $ & $ -0.046 \pm  0.001 $\\
& Radio$^{[3]}$  & $  0.62\pm   0.03  $     & $ -17.2 \pm   0.8 $     &    --                 &     --           \\
\hline
\end{tabular}
\end{center}
[1] The changing rates of the parameters from the joint PCA and HEXTE data.
[2] The changing rates of the parameters from the faked profiles that represent the aging effect of PCA.
[3] The radio results by \citep{Lyne et al.(2013)}.
\end{table*}

\begin{table*}
\caption {The intercepts of the linear fitting to the profile parameters}
\scriptsize{}
\footnotesize
\label{table:para_shift}
\begin{center}
\begin{tabular}{ l l c c c c }
\hline
\hline
& Instrument   & $\Phi$     & $R_{f}$   & $W_{1}$    & $W_{2}$     \\
&              & (\,\textordmasculine) &         & (\,\textordmasculine) & (\,\textordmasculine)   \\
\hline
& PCA           & $ 144.10 \pm   0.06 $ & $ 1.2330 \pm 0.0012 $ & $  14.36 \pm   0.05 $ & $  29.00 \pm   0.12 $  \\
& HEXTE      & $ 143.98 \pm   0.08 $ & $ 1.5058 \pm 0.0038 $ & $  14.85 \pm   0.10 $ & $  29.62 \pm   0.20 $  \\
\hline
\end{tabular}
\end{center}
\end{table*}

\end{document}